\newif
\ifconfver
\confvertrue        
\ifconfver
    \documentclass[10pt,journal,final,twoside]{IEEEtran}
\else
    \documentclass[11pt,draftcls,onecolumn]{IEEEtran}
\fi
\usepackage{xspace,empheq,fancybox,amsmath,amssymb,graphicx,epstopdf,epsfig,subfigure,syntonly,times,amsthm} \usepackage{psfrag,color,bm,array,cite}
\usepackage{url,cite,footnote,xspace,syntonly,algorithm,algpseudocode}
\usepackage{verbatim,multirow,slashbox}
\usepackage[T1]{fontenc}

\newtheorem{remark}{\bfseries Remark}
\input{mysymbol.sty}

\newcommand{\diag}{\mathrm{diag}}

\newcommand{\T}{\mathsf{T}}
\newcommand{\rmj}{\mathrm{j}}
\DeclareMathOperator*{\argmin}{argmin}
\begin{document}


\title{Data-driven Modeling for Distribution Grids \\ Under Partial Observability}
\author{
\IEEEauthorblockN{Shanny Lin},~\IEEEmembership{Graduate Student Member, IEEE}, and \IEEEauthorblockN{Hao Zhu},~\IEEEmembership{Senior Member, IEEE}

\thanks{\protect\rule{0pt}{3mm} This work has been supported by NSF Grants 1802319 and 2130706.} \thanks{\protect\rule{0pt}{3mm} The authors are with the Department of Electrical \& Computer Engineering, The University of Texas at Austin, Austin, TX, 78712, USA; Emails: {\{shannylin, haozhu\}{@}utexas.edu}.}
}

\renewcommand{\thepage}{}
\maketitle
\pagenumbering{arabic}

%
\begin{abstract} 
Accurately modeling power distribution grids is crucial for designing effective monitoring and decision making algorithms. This paper addresses the partial observability issue of data-driven distribution modeling in order to improve the accuracy of line parameter estimation. Inspired by the sparse changes in residential loads, we advocate to regularize the group sparsity of the unobservable injections in a bi-linear estimation problem.  
The alternating minimization scheme of guaranteed convergence is proposed to take advantage of convex sub-problems with efficient solutions. Numerical results using real-world load data on the single-phase equivalent of the IEEE 123-bus test case have demonstrated the accuracy improvements of the proposed solution over existing work for both parameter estimation and voltage modeling.
\end{abstract}



\section{Introduction} \label{sec:Intro}

Power distribution grids have witnessed a rapid integration of distributed energy resources (DERs) such as solar photovoltaic (PV) generation, plug-in electric vehicles, and energy storage. Highly variable and random DER outputs lead to growing concerns on the operational guarantees and reliability of distribution feeders. To attain effective monitoring and management for system-wide DERs, utility operators need the accurate modeling of feeder networks \cite{ardakanian2019identification}. 
Unlike transmission systems,  distribution grids typically have only partial or unreliable knowledge of its network topology and line parameters \cite{liao2018urban}. 
Thus, improving the modeling accuracy is of great importance to adapt to an increasingly dynamic operational paradigm for distribution grids. 




The rising deployment of advanced sensing and communication technologies \cite{morello2017advances,saldana2020advanced} 
has given rise to unprecedented opportunities for data-driven modeling of distribution grids. The nodal DERs are typically equipped with high-rate sensors at their inverter interfaces, while feeder-level measurements have greatly increased as well thanks to distribution phasor measurement units (D-PMUs) or other power line sensors. 
Existing data-driven approaches can be broadly categorized into two groups according to the availability of topology. Assuming that the topology is known, the first group \cite{xu2019data,xu2019datadriven,nowak2020measurement,li2021distribution} has advocated to use fast-sampled nodal voltage and injection data to identify the line parameters through linear regression. The other group \cite{liao2018urban,yu2018patopaem,zhao2017learning,ardakanian2019identification} considers the identification of unknown topology, which may be common for secondary feeders, by utilizing the  sparse connectivity of tree-topology feeders. Regardless of the topology knowledge, line parameter estimation is a fundamental problem in distribution modeling and important to consider in this sense. 

One key challenge faced by distribution modeling is the status quo of \textit{partial observability}. Limited by the deployment cost, high-rate sensors are only available at a subset of nodes, with the other nodes \textit{unobservable}. Recognizing this issue, \cite{nowak2020measurement,li2021distribution} simply neglects the unobserved nodes by assuming their power injections are slowly varying. Nonetheless, this assumption fails to hold for residential customer loads which consist of \textit{sudden, large} changes due to the household activities; see e.g., \cite{lin2021enhancing}. Other approaches \cite{sharon2012topology,cavraro2019real,cavraro2018graph,deka2020joint} address the partial observability issue by using the second-order statistics of nodal voltage and injection, which do not contain the dynamics from fast data samples. Thus, it still remains challenging to accurately estimate the line parameters from fast data streams to account for the partial observability.

The goal of this paper is to develop an effective line parameter estimation algorithm that can directly incorporate the load characteristics of unobservable nodes. The consecutive differences of voltage and injection measurements are related to the unknown parameters based on the linearized distribution flow model. This leads to a linear regression problem by assuming full observability and using fixed r-to-x ratios to address some numerical conditioning issues. 
To account for the unobservable nodes, we propose to include their injections as unknowns to solve a bi-linear estimation problem. Inspired by the fact that residential loads have \textit{sparse changes}, we advocate to regularize the unknown injections using the group L1-norm on both the nodal active and reactive power differences. 
An alternating minimization (AM) algorithm is developed to solve the resultant bi-linear problem by alternatingly updating the unknown injections and line parameters, both of which are convex sub-problems. As the iterative objective cost is non-increasing during these updates, the proposed AM algorithm is guaranteed to converge. To sum up, the main contribution of the present work lies in the direct consideration of partially observed feeders and the utilization of a meaningful regularization term to improve the accuracy of line parameter estimation. Our proposed solution enjoys computation efficiency and guaranteed convergence.

The rest of this paper is organized as follows. Section \ref{sec:SysModel} presents the system modeling to relate the changes of voltage and injection data to the unknown line parameters. In Section \ref{sec:Formulation}, we formulate the bi-linear problem by regularizing the unknown injections and solve it using the alternating minimization updates. Numerical results using real-world load data on the single-phase equivalent of the IEEE 123-bus test case are presented in Section \ref{sec:Results} to demonstrate the performance improvements of the proposed solution in terms of estimation accuracy and voltage modeling. The paper is wrapped up in Section \ref{sec:Conclusion}. 


\textit{Notation:} Upper (lower) boldface symbols stand for matrices (vectors); $(\cdot)^{\T}$ stands for matrix transposition; $\|\cdot\|_G$ the group L1-norm; $\|\cdot\|_p$ the vector p-norm for $p\geq 1$; $\bbI$ $(\mathbf 1)$ stands for the identity matrix (all-one vector) of appropriate size; $\diag(\cdot)$ represents the diagonal matrix; upper calligraphic symbols denotes sets; symbols with $\tilde{\cdot}$ indicate the difference between two consecutive time instances; and $(a+ \rmj b)$ represents a complex quantity.
%
\section{System Modeling} \label{sec:SysModel}

Consider a radial distribution feeder consisting of $(N+1)$ buses and $L$  line segments that can be represented by a graph $\mathcal G = (\mathcal N^+, \mathcal L)$. As distribution grids are typically of tree topology, the number of lines $L = (N+1)-1 = N$. The set of buses $\mathcal N^+ := \{0,1,\cdots,N\}$ include the reference bus 0, and the remaining $N$ PQ buses in $\ccalN:=  \mathcal N^+{\setminus \{0\}}$. Similarly, $\mathcal L = \{1,\cdots,L\}$ stands for the set of lines. Per bus $n$, let $v_n$ and $(p_n + \rmj q_n)$ denote the voltage magnitude and complex power injection in per unit (p.u.), respectively. Without loss of generality (Wlog), voltage magnitude will be referred to as voltage for the rest of this paper.  Collecting all nodal variables of non-reference buses forms vectors $\bbv$, $\bbp$, and $\bbq$, all of length $N$.  For each line $\ell$, let $(r_\ell + \rmj x_\ell)$ denote the line impedance in p.u. with all line resistances and reactances respectively collected in vectors $\bbr$ and $\bbx$, both of length $L$. For simplicity, this paper considers only single-phase feeder models, as in most earlier work {\cite{xu2019datadriven,liao2018urban,yu2018patopaem,zhao2017learning,deka2020joint,cavraro2019real,cavraro2017voltage,cavraro2018graph}}. 
Extensions to multi-phase feeders are possible by using the multi-phase linearized models in e.g., \cite{gan2014convex,liu2017distributed,liu2018hybrid}. 

Given the line parameters, the ac power flow asserts a nonlinear model of the voltage $\bbv$ with respect to (wrt) the injections $(\bbp,~\bbq)$, making the problem of line parameter estimation more complicated. Thus, we adopt the linearized DistFlow (LDF) model \cite{baran1989network}, which has been popularly used for distribution grid modeling and analysis. The LDF model uses \textit{graph-based matrices} to capture the topology of distribution feeders. Let $\check{\bbM} \in \{0, \pm 1\}^{(N+1)\times L}$ represent the node-edge incidence matrix of graph $\mathcal G$, with its submatrix $\bbM \in \mathbb R^{N\times L}$ by removing the row corresponding to bus 0. Recalling that $L=N$ for tree-topology feeders, we have an invertible square matrix $\bbM$ for connected graph $\ccalG$. Given the reference voltage $v_0$ which is typically 1, the LDF approximation yields that
\begin{align} \label{eqn:LDF}
\bbv \approxeq \bbR (\bbr) \bbp + \bbX (\bbx) \bbq + \mathbf 1 v_0
\end{align}
where  matrices $\bbR(\bbr), ~\bbX(\bbx) \in \mathbb R^{N\times N}$ represent the network effects of respectively resistance $\bbr$ and reactance $\bbx$, defined as
\begin{subequations} \label{eq:RX}
\begin{align}
\bbR(\bbr) &:= \bbM^{- \T} \diag(\bbr) \bbM^{-1} \\
\bbX(\bbx) &:= \bbM^{- \T} \diag(\bbx) \bbM^{-1}
\end{align}
\end{subequations}
where $\diag(\cdot)$ returns a diagonal matrix with the input vector as its diagonal entries. Clearly, they capture the network-wide sensitivity of voltage wrt the injected powers for fully unloaded feeders (i.e., $\bbp =  \bbq = \mathbf 0$). Although the incidence matrix $\bbM$ is very sparse, its inverse is typically a full matrix and so are $\bbR$ and $\bbX$.

As the LDF model can be viewed as the sensitivity analysis for unloaded feeders, its accuracy is improved by modeling the voltage difference due to changes in the power injections, instead of modeling the actual voltage. To this end, let $t  \in \{ 1, \ldots, T\}$ index the discrete-time samples of the feeder-wide variables.  By applying \eqref{eqn:LDF} over two consecutive time instances, we obtain the voltage difference $\tilde{\bbv}_t \coloneqq \bbv_t - \bbv_{t-1}$ as
\begin{align} \label{eqn:diff_model}
\tilde{\bbv}_t &\approxeq \bbR (\bbr) \tilde{\bbp}_t + \bbX (\bbx) \tilde{\bbq}_t \nonumber\\
& \coloneqq \bbA(\tilde{\bbs}_t) \bbtheta
\end{align} 
where vectors $\tdbbp_t$ and $\tdbbq_t$ are similarly defined as the differences of consecutive injections. To relate $\tdbbv_t$ with the unknown line parameters in vector $\bbtheta \coloneqq \left[ \bbr;~\bbx \right] \in \mathbb R^{2L}$, we form the matrix function $\bbA(\tilde{\bbs}_t) \in \mathbb R^{N\times 2L}$ which linearly depends on the total power difference $\tilde{\bbs}_t \coloneqq \left[ \tilde{\bbp}_t; \tilde{\bbq}_t \right]$. The voltage difference in \eqref{eqn:diff_model} is approximately linear wrt the line parameters. Hence, assuming all buses are metered with high-rate sensors, the line parameters can be easily estimated by solving a linear regression problem as given by
\begin{align} \label{eqn:lin_regression}
\min_{\bbtheta} ~ \frac{1}{T} \sum_{t=1}^T \left\| \tilde{\bbv}_t - \bbA(\tilde{\bbs}_t) \bbtheta \right\|_2^2
\end{align}
that minimizes the model fitting error for \eqref{eqn:diff_model}. This is an ordinary least-squares (OLS) problem and $\bbtheta$ can be directly solved either using the closed-form solution or through gradient descent updates. Nonetheless, under the limited observability in current distribution grids, only a subset of buses in $\mathcal N$ 
are equipped with high-rate meters to provide the difference data needed for solving \eqref{eqn:lin_regression}. To account for this \textit{partial observability} issue, the ensuing section will solve it as a bi-linear problem by assuming non-frequent changes in power injections. Before that, we provide two remarks here. 

\begin{remark}[Modeling accuracy] Although other linearized models may obtain higher accuracy {\cite{bernstein2018load}}, we have selected the LDF model for its simple relation to the line parameters of interest. Due to the dependence on $\bbM$ in \eqref{eq:RX}, entries of matrices $\bbR$ and $\bbX$ capture the accumulated resistance and reactance, respectively, along the common path \cite{cavraro2019real}. 
In addition, the LDF model in \eqref{eqn:LDF} can be viewed as the sensitivity analysis  for small $\bbp$ and $\bbq$. Thus, its modeling accuracy can be enhanced by considering the consecutive differences as in \eqref{eqn:diff_model}. Interestingly, even if the actual sensitivity matrices slightly differ from $\bbR$ and $\bbX$,  the data-driven linear regression in \eqref{eqn:lin_regression} would still produce the best linear representation under small perturbation. Therefore, the LDF model is an excellent choice for data-driven distribution modeling.  
\end{remark}

\begin{remark}[Numerical issues] \label{rmk:numerical} Two potential issues may affect the numerical stability of the linear regression in \eqref{eqn:lin_regression}. First, if the changes $\tilde{\bbp}_t$ and $\tilde{\bbq}_t$ are synchronized due to e.g., similar power factor, then columns of matrix $\bbA(\tilde{\bbs}_t)$ would be highly correlated. This could affect the identifiability of \eqref{eqn:lin_regression} in terms of distinguishing $\bbr$ and $\bbx$.  One solution is to fix the r-to-x ratio $\bbz = \{z_\ell\} = \{r_\ell / x_\ell\}$ for each line such that $\bbr = \diag(\bbz)\bbx$. This information can be obtained based on the conductor datasheet and the line configuration \cite[Ch.~4]{kersting2006distribution}. With $\bbz$ given, the problem becomes to estimate either resistance $\bbr$ or reactance $\bbx$, and our numerical studies later on will perform the reactance estimation only. 
Second, the incidence matrice $\bbM$ is prone to have very large condition number, which can affect the estimation in \eqref{eqn:lin_regression}. For general graph incidence, spectral analysis results in \cite{frangioni2001spectral} show that its smallest singular value could follow $\sigma_{\min} \sim \ccalO(1/N)$ while the largest one follows $\sigma_{max} \sim \ccalO(\sqrt{N})$. Therefore, the condition number of the regression matrix $\bbA(\tdbbs_t)$ may grow with large $N$ at at least $\ccalO(N^{3/2})$. This numerical issue can be mitigated by introducing a regularization of $\bbtheta$, as discussed soon. 
\end{remark}


\section{Distribution Modeling under Partial Observability} \label{sec:Formulation}

We partition the set of buses $\ccalN$ into two subsets $\mathcal O$ and $\ccalU$, which respectively collects all the observable and unobservable buses. To form the differences in \eqref{eqn:diff_model}, buses in $\ccalO$ need to have access to high-rate meters 
such as D-PMUs or inverter sensors. Note that if buses only have smart meters, the slow sampling rates may not be sufficient for providing the data in \eqref{eqn:diff_model}, and thus they will be in $\ccalU$.
This bus partition leads to a similar partition of $\tilde{\bbv}_t = [\tilde{\bbv}_t^\mathcal O;~\tilde{\bbv}_t^\mathcal U]$ and $\tilde{\bbs}_t = [\tilde{\bbs}_t^\mathcal O; ~ \tilde{\bbs}_t^\mathcal U]$. Under the LDF model \eqref{eqn:diff_model}, the voltage depends on the full vector of $\tilde{\bbs}_t$, such that with only $\tilde{\bbs}_t^\ccalO$ available, the modeling problem in \eqref{eqn:lin_regression} becomes
\begin{align} \label{eqn:bilinear_prob}
\min_{\bbtheta, \{\tilde{\bbs}_t^\mathcal U\}} ~ \frac{1}{T} \sum_{t=1}^T \left\| \tilde{\bbv}_t^\mathcal O - \bbA (\tilde{\bbs}_t^\mathcal O) \bbtheta - \bbA (\tilde{\bbs}_t^\mathcal U) \bbtheta \right\|_2^2
\end{align}
where the unknown $\tilde{\bbs}_t^\mathcal U$ is treated as an optimization variable. This becomes a bi-linear regression problem with a high degree of identifiability and model overfitting issues as the regression matrix $\bbA (\tilde{\bbs}_t^\mathcal U)$ changes across time $t$. The following approach has been developed to solve the problem \eqref{eqn:bilinear_prob}. 

The zero-injection (ZI) approach in \cite{nowak2020measurement,li2021distribution} 
assumes that non-metered buses have slowly changing loads 
as 
high-rate meters are usually installed for loads of high variability. Accordingly, the problem \eqref{eqn:bilinear_prob} can use $\tilde{\bbs}_t^\mathcal U \approxeq \mathbf 0$ to obtain $\bbA (\tilde{\bbs}_t^\mathcal U) \bbtheta = \mathbf 0$ which again leads to a linear regression problem. Although this approach results in an easily solvable modeling problem, it fails to represent the actual load characteristics as detailed soon. Hence, its effectiveness can greatly decrease with a large percentage of unobservable buses.
	
%


Unlike transmission-level loads, residential loads are known to have sudden changes due to large appliance activities or other DERs; see e.g., \cite{lin2021enhancing}. These changes are mostly \textit{infrequent}, but can cause large variations of load profiles. As a result, these load variations cannot be simply assumed to be always zero, as done by the ZI approach. Interestingly, this {infrequent} feature of load variations motivates a meaningful regularization term for the unknown $\tilde{\bbs}_t^\mathcal U$. 

Infrequent load changes manifest in a high likelihood of the corresponding entries of $\tilde{\bbp}_t^\mathcal U$ and $\tilde{\bbq}_t^\mathcal U$ being zero, or {very sparse}. Furthermore, each pair 
$\{\tilde p_{n,t},\tilde q_{n,t} \}_{n\in \mathcal U}$ is \textit{jointly sparse} as the activity of residential appliances or DERs leads to synchronized changes of both active and reactive power demand. To promote this joint sparsity, we propose to regularize \eqref{eqn:bilinear_prob} using the group-L1 norm of $\tdbbs_t^\mathcal U$ as defined by
\begin{align}
\|\tilde{\bbs}_t^\mathcal U \|_G := \sum_{n \in \mathcal U} \left \| [\tilde p_{n,t}^\mathcal U, ~\tilde q_{n,t}^\mathcal U] \right \|_2,~\forall t. \label{eqn:gl1}
\end{align}
This regularization can be thought of as the L1-norm for the jointly sparse pairs, and thus is powerful for promoting sparsity at the group level \cite{yuan2006glasso}. 
The group-L1 norm  \eqref{eqn:gl1} is strongly convex as the summation of L2-norms, and thus can lead to tractable solutions. In addition, we regularize the line parameters using $\|\bbtheta\|_2^2$, due to the aforementioned numerical conditioning issue. This is a popular regularization choice for general predictive modeling tasks to mitigate any possible model overfitting; see e.g., {\cite[Ch.~6]{james2013introduction}.} 
If a nominal value for $\bbtheta$ is known as $\bbarbbtheta$ based on e.g., conductor datasheet or historical data,  we can also regularize on the deviation using $\|\bbtheta-\bbarbbtheta\|_2^2$. Together with the two regularization terms, we develop the following formulation for estimating the line parameters: 
\begin{align} \label{eqn:reg_bilinear_prob}
\min_{\bbtheta, \{\tilde{\bbs}_t^\mathcal U\}} ~  \frac{1}{T} \sum_{t=1}^T  &\left[ \left\| \tilde{\bbv}_t^\mathcal O - \bbA (\tilde{\bbs}_t^\mathcal O) \bbtheta - \bbA (\tilde{\bbs}_t^\mathcal U) \bbtheta \right\|_2^2 + \lambda \|\tilde{\bbs}_t^\mathcal U \|_G \right] \nonumber \\
&~~~+ \alpha  \|\bbtheta \|_2^2
\end{align}
where $\lambda, \alpha > 0$ are the regularization coefficients or hyperparameters. They are typically tuned using cross-validation \cite[Ch.~5]{james2013introduction}. This method determines the best pair of $(\lambda, \alpha)$ based on how the model obtained by solving  \eqref{eqn:reg_bilinear_prob} on the training dataset performs on the test dataset that the model has not seen yet. For example, a $K$-fold cross validation randomly splits the full dataset into approximately $K$ equal groups. The model is trained on $(K-1)$ groups and validated on the remaining one by computing the model accuracy such as the mean squared error. This procedure is repeated $K$ times with a different group used to validate the model each time. This results in $K$ accuracy estimates and 
the average accuracy for the given $(\lambda, \alpha)$ pair is used for selecting the best hyperparameter values. We will employ this method for determining the hyperparameter settings in our numerical tests.

{
\begin{remark}[Additional measurements] The recovery problem \eqref{eqn:reg_bilinear_prob} can incorporate other types of feeder-level measurements to improve the estimation accuracy. For example, the aggregated power demand at the reference bus is approximately equal to the total power injection $\bbs_t$, and thus we can use its measurements  to pose constraints or regularization terms on $\tilde{\bbs}_t^\mathcal U$. Other line power/current flow measurements and $\bbtheta$ being non-negative can be considered similarly. We develop the solution method for the basic formulation in \eqref{eqn:bilinear_prob}, which can be modified to account for those measurements too. 
\end{remark}
}


The problem \eqref{eqn:reg_bilinear_prob} is non-convex and difficult to solve due to the bi-linear relation between $\bbtheta$ and $\tilde{\bbs}_t^\mathcal U$. Nonetheless, by fixing either group of unknowns, this bi-linear problem becomes convex wrt the other group and can be solved efficiently. Hence, we develop an alternating minimization (AM) based scheme to solve \eqref{eqn:reg_bilinear_prob} by iteratively updating the two groups of variables. The proposed AM solution works with the initial parameter values $\bbtheta[0]$ at iteration $i = 0$. This can be set as the nominal values or using the ZI approach as discussed earlier.  With $\bbtheta[i-1]$ given per iteration $i \geq 1$,  we have two sub-problems, each for updating $\{\tdbbs_t^\ccalU[i]\}$ and $\bbtheta [i]$ alternatively, as given by
\begin{subequations} \label{eqn:subprob}
\begin{align} 
\tilde{\bbs}_t^\mathcal U[i] &= \argmin_{\tilde{\bbs}_t^\mathcal U} ~ \left\| \tilde{\bbv}_t^\mathcal O - \bbA (\tilde{\bbs}_t^\mathcal O) \bbtheta[i-1] - \bbA (\tilde{\bbs}_t^\mathcal U) \bbtheta[i-1] \right\|_2^2 \nonumber \\
&~~~+ \lambda \|\tilde{\bbs}_t^\mathcal U \|_G, ~ \forall t=1,\cdots,T \label{eqn:altmin_su}\\
\bbtheta[i] &= \argmin_{\bbtheta} ~ \frac{1}{T} \left[ \sum_{t=1}^T \left\| \tilde{\bbv}_t^\mathcal O - \bbA (\tilde{\bbs}_t^\mathcal O) \bbtheta - \bbA (\tilde{\bbs}_t^\mathcal U [i]) \bbtheta \right\|_2^2 \right] \nonumber \\
&~~~ + \alpha  \|\bbtheta \|_2^2. \label{eqn:altmin_theta}
\end{align}
\end{subequations}
With given $\bbtheta[i-1]$, we can update $\{\tdbbs_t^\ccalU[i]\}$ as a linear regression problem regularized by the group-L1 norm. Interestingly, this sub-problem is decoupled into $T$ separate updates, each for  $\tilde{\bbs}_t^\mathcal U$ in \eqref{eqn:altmin_su}. This is a group-Lasso problem which can be solved using generic convex conic optimization solvers such as MOSEK \cite{mosek}. 
The hyperparameter $\lambda$ here is intuitively a shrinkage penalty that controls the sparsity level of the pairs $\{[\tilde p_{n,t},\tilde q_{n,t}] \}_{n\in \mathcal U}$. When $\lambda$ value increases, the group-L1 norm will force more pairs to become all zeros, i.e. by setting $[\tdp_{n,t},\tdq_{n,t}] = [0,0]$. Note that $\lambda$ scales with the number of unobserved nodes to balance between the observed voltage fitting error and unobserved injection regularization. 
One can also develop efficient updates for \eqref{eqn:altmin_su} using  block coordinate descent 
\cite{yuan2006glasso}, which alternatively optimizes each pair $[\tdp_{n,t},\tdq_{n,t}]$ while fixing the other pairs. This efficient solution is useful for extending to online adaptive solutions in future.


Meanwhile, the $\bbtheta$ sub-problem \eqref{eqn:altmin_theta} is a simple ridge regression which has a closed-form solution given by 
\begin{align} \label{eqn:theta_closedform}
\bbtheta[i] = (\bbG [i])^{-1} \bbh [i]
\end{align} 
where
\begin{align*}
\bbG [i] &= \frac{1}{T} \sum_{t=1}^T \left[  \left( \bbA (\tilde{\bbs}_t^\mathcal O) + \bbA (\tilde{\bbs}_t^\mathcal U [i]) \right)^\T \left( \bbA (\tilde{\bbs}_t^\mathcal O) + \bbA (\tilde{\bbs}_t^\mathcal U [i]) \right) \right] \\
&~~~~~~~ + \alpha \bbI, \\
\bbh [i] &=  \frac{1}{T} \sum_{t=1}^T \left[ \left( \bbA (\tilde{\bbs}_t^\mathcal O) + \bbA (\tilde{\bbs}_t^\mathcal U [i]) \right)^\T \tilde{\bbv}_t^\mathcal O \right]. 
\end{align*}
The hyperparameter $\alpha > 0$ here controls the L2-norm of $\bbtheta$. 
Similarly,  large $\alpha$ values would lead entries of $\bbtheta$ to be small. Notice that the condition $\alpha >0$ ensures the symmetric matrix $\bbG [i]$ to be invertible with all positive eigenvalues. Hence, the regularization $\|\bbtheta\|_2^2$ is helpful for improving the numerical condition of solving \eqref{eqn:reg_bilinear_prob}. 

The detailed steps  of the AM scheme are tabulated in Algorithm \ref{alg:altmin}. Note that the hyperparameters $(\lambda,\alpha)$ can be tuned up based on the aforementioned cross-validation procedure. We initialize the iterations by setting  $\tilde{\bbs}_t^{\mathcal U} [0] = \mathbf 0$ for all $t$. Under this initialization, the estimated line parameters $\bbtheta[0]$ from \eqref{eqn:theta_closedform} is equivalent to the solution based on the ZI approach. 
As the sub-problems in \eqref{eqn:subprob} are unconstrained convex problems, each update is guaranteed to produce the feasible global optimum that can reduce the objective value of \eqref{eqn:reg_bilinear_prob}. Hence, the convergence of Algorithm \ref{alg:altmin} can be established as in Proposition \ref{prop:cong}.

\begin{algorithm}[t]
\caption{Alternating Minimization (AM) for Solving \eqref{eqn:reg_bilinear_prob}}
\begin{algorithmic}[1]
\State \textbf{Input:} Observed data $\{ \tilde{\bbv}_t^\mathcal O,~\tilde{\bbp}_t^\mathcal O,\tilde{\bbq}_t^\mathcal O\}_{t=1}^T$, the hyperparameters $(\lambda,\alpha)$, and an iteration stopping threshold $\epsilon$.
\State \textbf{Output:} Estimated line parameters $\bbtheta$ and unobserved power injections $\{ \tilde{\bbs}_t^\mathcal U \}_{t=1}^T$.
\State \textbf{Initialize}: Set the unobserved data  $\{ \tilde{\bbs}_t^\mathcal U[0] \}_{t=1}^T \leftarrow \mathbf 0$. 

\State Update $\bbtheta[0]$ as the ZI solution using \eqref{eqn:theta_closedform}. 
\State Set  the iteration number $i=0$.

\While{$\|\bbtheta[i] -\bbtheta[i-1]\|_2 \geq \epsilon $}
\State Update $i\leftarrow i+1$.
\State Update $\tilde{\bbs}_t^\mathcal U [i]$ using \eqref{eqn:altmin_su} for all $t=1,\ldots, T$.
\State Update $\bbtheta [i]$ using \eqref{eqn:theta_closedform}.
\EndWhile
\State \textbf{Return:} $\bbtheta[i],~\{ \tilde{\bbs}_t^\mathcal U [i]\}_{t=1}^T$
\end{algorithmic}
\label{alg:altmin}
\end{algorithm}

\begin{proposition}[Convergence of AM \cite{tseng2009block}] At each iteration, the convex sub-problems in \eqref{eqn:subprob} can attain the feasible  
globally optimal solution. Accordingly, the objective value in \eqref{eqn:reg_bilinear_prob} is non-increasing as the iterations continue, and thus Algorithm \ref{alg:altmin} converges to a stationary point of \eqref{eqn:reg_bilinear_prob}. 
	\label{prop:cong}
\end{proposition}

\begin{remark}[Uniqueness of recovery] 
By simplifying the sparsity modeling, the bi-linear estimation can be viewed as a sparse dictionary learning problem of recovering $\bbD$ and $\bbY$ from their product in general \cite{aharon2006uniqueness}. The dictionary $\bbD$ here depends on $\bbtheta$ according to the matrices in \eqref{eqn:LDF}, while the sparse coding matrix $\bbY$ represents the unknown injection differences. The uniqueness of recovery for dictionary learning requires several conditions. First, as a full-rank matrix, it is ideal for $\bbD$ to have low \textit{mutual coherency}, or weakly correlated columns. Second, the columns of $\bbY$ need to be sparse, with bounded number of nonzero entries. Third, to ensure the overall system contains sufficient information, they should consist of all possible combinations of sparse support vectors. The latter two conditions require the unknown injection differences to ideally have infrequent occurrence and random magnitude, which generally hold for residential load changes. This observation inspires us to formally analyze the recovery guarantees.
\end{remark}

\section{Numerical Results} \label{sec:Results}

We evaluate the effectiveness of Algorithm \ref{alg:altmin} using real-world load data. The feeder model is based on the single-phase equivalent of the IEEE 123-bus test case \cite{ieee123}. 
There are a total of 90 load nodes with buses 66, 85, 96, 114, 151, and 250 having PV generation. The active power load profiles $\{\bbp_t\}_{t=1}^T$ of minute-level resolution were obtained from the PecanStreet Dataport \cite{pecan} for homes located in Austin, TX. The reactive power ones were synthetically generated using a random power factor in the range $[0.9,~0.95]$. Based on these profiles, the actual voltage maganitudes $\{\bbv_t\}_{t=1}^T$ were solved using the fixed-point ac power flow solver as developed in \cite{bernstein2018load}.

We have considered three scenarios of partial observability,  by observing the data from 45\%, 
60\%, or 75\% of the load nodes. Under the 45\% observability scenario, high-rate meters are placed at all leaf nodes (end-buses) only. The 
60\% and 75\% scenarios have additional meters placed along the feeder laterals. For all three scenarios, we have observed the voltage at four critical feeder locations, buses 13, 18, 57, and 67. The full (100\%) observability scenario with all nodes metered is also provided as a baseline. Algorithm \ref{alg:altmin} was implemented in the MATLAB\textsuperscript{\textregistered} R2020b simulator on a desktop with Intel\textsuperscript{\textregistered} Core\texttrademark~i5 CPU @ 3.40 GHz and 16 GB of RAM. The convex sub-problem \eqref{eqn:altmin_su} was solved using the CVX toolbox \cite{cvx} with the MOSEK solver.

\begin{figure}[tb!]
	{\footnotesize
		\centering
		\includegraphics[width=0.85\linewidth]{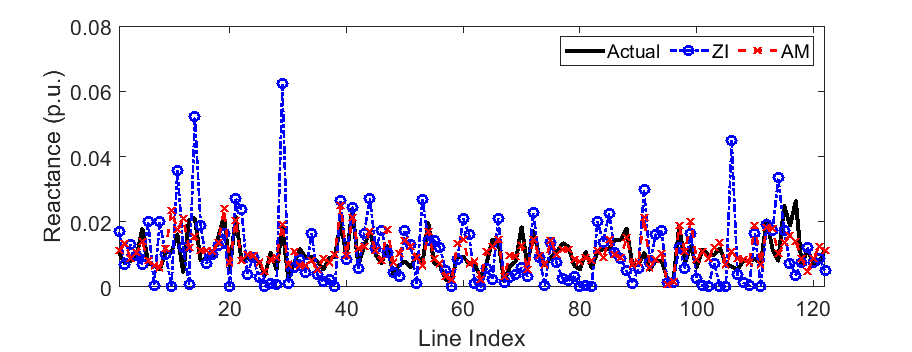}
		\centerline{(a)}
		\includegraphics[width=0.85\linewidth]{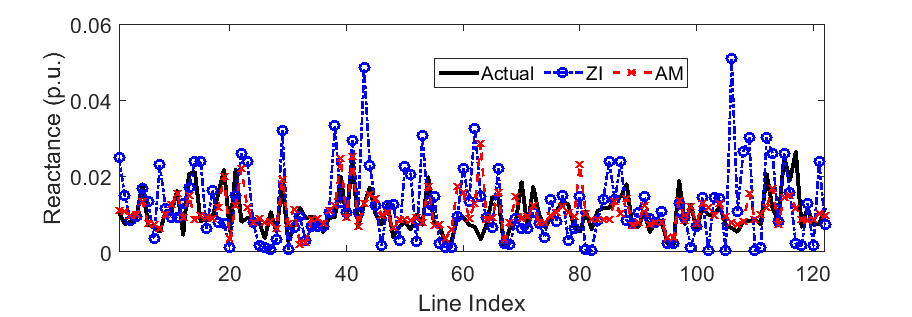}
		\centerline{(b)}
		\includegraphics[width=0.85\linewidth]{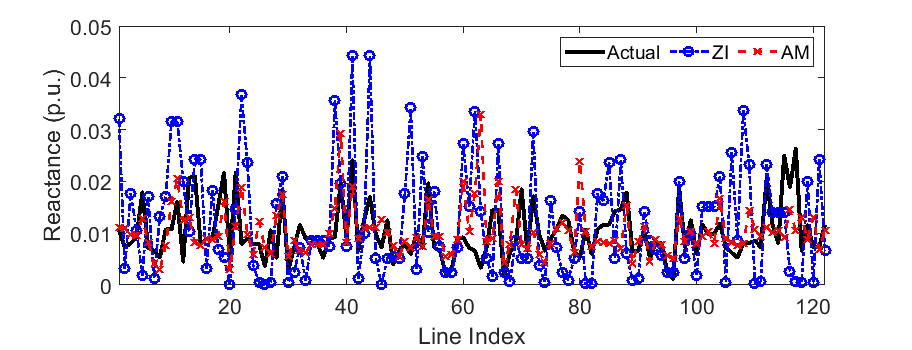}
		\centerline{(c)}
		\caption{Estimated line reactances $\hat{\bbx}$ by the proposed Algorithm \ref{alg:altmin} (AM) and ZI method in comparison with the actual values under (a) 75\%, (b) 60\%, and (c) 45\% observability.}
		\label{fig:x}}
\end{figure}

We have used the proposed Algorithm \ref{alg:altmin} to fit one day of load data, or 1440 samples. As mentioned in Remark \ref{rmk:numerical}, we assume the r-to-x ratios are known and only estimate the line reactances 
$\bbx$. To determine the hyperparameters $\alpha$ and $\lambda$, we used $5$-fold cross validation based on the one-day dataset. The value of $\alpha = 5e{-7}$ was identified as the best choice for all three scenarios. Meanwhile, the best choice of $\lambda$ value for the 45\%, 60\%, and 75\% scenarios were respectively identified as  $5e{-9}$, $1e{-8}$, and $5e{-4}$. Recall that $\lambda$ scales with the number of unobserved nodes, and thus varies under different observability levels. 
Algorithm \ref{alg:altmin} has used these hyperparameters to obtain the estimated $\hat{\bbx}$. Fig.~\ref{fig:x} compares the estimated $\hat{\bbx}$ obtained by the proposed AM method (red) and the initial ZI solution (blue) with the actual values (black), for all three partial observability scenarios. The ZI initialization incurs very  large estimation error at certain lines (almost doubling the reactance values), while the proposed Algorithm \ref{alg:altmin} is able to maintain a consistent error accuracy throughout the feeder. Moreover, the resulting normalized total vector error (TVE) given by $\| \hat{\bbx} - \bbx \|_2 / \| \bbx \|_2$ is tabulated in Table \ref{tbl:TVE}, which also includes the full observability baseline. Clearly, the TVE increases as the number of measurement locations reduces for both methods. Compared to the proposed method, ZI initialization has high error percentage ($\sim 90\%$) under partial observability. Notably, our proposed method under 75\% observability attains almost the same accuracy as the baseline of 100\% observability, while still maintaining satisfactory performance with lower observability. Therefore, by accounting for the sparse load changes, the proposed Algorithm \ref{alg:altmin} greatly improves the ZI solution by providing a consistent parameter estimation throughout the feeder.

\begin{table}[tb!]
	\centering 
	\caption{Total vector error (TVE) of reactance estimation.}
	\begin{tabular}{ |c|c|c| } 
		\hline
		Observability (\%) & AM (\%) & ZI (\%) \\
		\hline
		100 & 35.52 & 35.52 \\ 
		75& 35.57 & 87.96 \\ 
		60& 42.89 & 88.38 \\ 
		45& 50.60 & 95.20 \\
		\hline
	\end{tabular}
	\label{tbl:TVE}
\end{table}

\begin{figure}
	{\footnotesize
		\centering
		\includegraphics[width=0.65\linewidth]{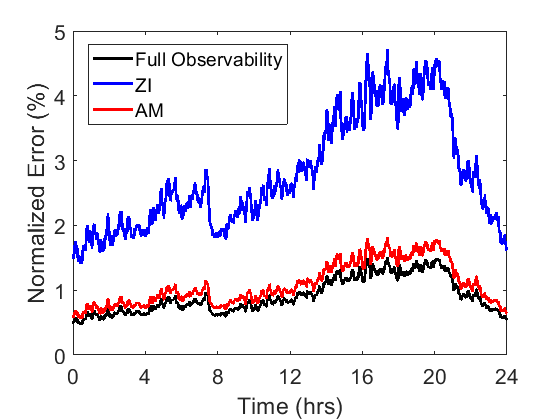}
		\centerline{(a)}
		\includegraphics[width=0.65\linewidth]{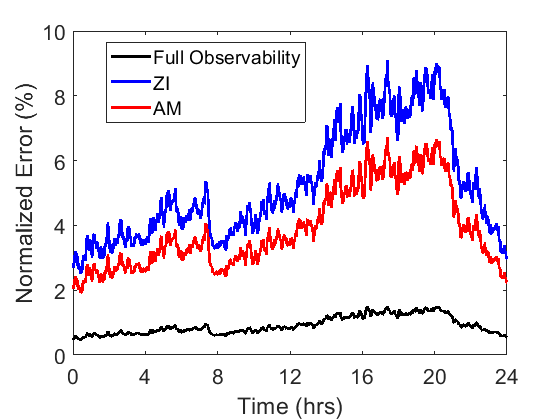}
		\centerline{(b)}
		\includegraphics[width=0.65\linewidth]{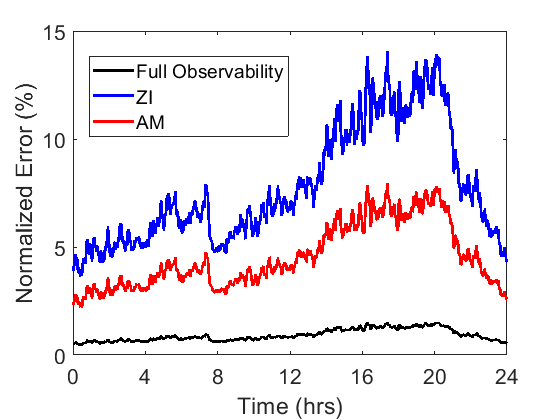}
		\centerline{(c)}
		\caption{Normalized voltage magnitude error of all load nodes over time for (a) 75\%, (b) 60\%, and (c) 45\% observability.}
		\label{fig:volt_L2_error}}
\end{figure}

We have further validated the estimated parameters $\hat{\bbtheta}$ using another day of load data, by evaluating its 
performance 
in voltage prediction on the new dataset. The voltage prediction errors were computed by the ac power flow model based on the 
estimated reactances. Fig.~\ref{fig:volt_L2_error} plots the normalized percentage error for the voltage vector across all load nodes under different observability levels. The voltage prediction performance of the proposed AM method under 75\% observability is very comparable to the baseline performance of full observability, which significantly reduces the error from the ZI initialization. The performance of both the proposed AM method and ZI solution  degrades with lower observability, but the former still maintains the prediction error to be around 5\%. 
By accounting for the unobservable nodes, our proposed Algorithm \ref{alg:altmin} can attain an improved estimation of line parameters and thus enhance the voltage prediction capability, for the problem of distribution modeling under partial observability.


\section{Conclusion and Future Work} \label{sec:Conclusion}

This paper developed a distribution modeling method under the practical consideration of partial observability. By accounting for the unobservable loads as sparse changes, the problem of estimating line parameters becomes a bi-linear optimization problem. Meaningful regularization terms on the unknown parameters and the unobservable injections have been introduced to address the numerical issues and lack of measurements. 
Using the existing 
zero-injection (ZI) approach as an initialization, we developed an alternating minimization (AM) method to update each one of the two groups of unknowns by fixing the other. The proposed AM method leads to efficient updates by solving convex sub-problems and can converge to a stationary point. Numerical tests using real-world load data on the single-phase equivalent of the IEEE 123-bus 
test case have demonstrated the effectiveness of the proposed method for improving the performance in both parameter estimation and voltage modeling.

Exciting research directions open up regarding the development of efficient adaptive updates and the uniqueness analysis for recovering the unknowns. Furthermore, we are interested to extend the current results to incorporate unknown r-to-x ratios and even unknown feeder topology. 

%
\bibliographystyle{IEEEtran}
\itemsep2pt
\bibliography{ref}

%
%

\end{document}